\documentstyle[aps,prl,epsf,floats,rotating]{revtex}

\newcommand{\nn}{\nonumber}
\newcommand{\beq}{\begin{equation}}
\newcommand{\eeq}{\end{equation}}
\newcommand{\bea}{\begin{eqnarray}}
\newcommand{\eea}{\end{eqnarray}}
\newcommand{\ben}{\begin{eqnarray*}}
\newcommand{\een}{\end{eqnarray*}}

\def\D0{D\O}

\begin{document}

\twocolumn[\hsize\textwidth\columnwidth\hsize\csname@twocolumnfalse\endcsname

\title{The Leading Particle Effect from Heavy-Quark Recombination}

\author{Eric Braaten and  Yu Jia }

\address{Physics Department, Ohio State University, Columbus,
OH 43210}

\author{Thomas Mehen }

\address{Department of Physics, Duke University, Durham NC 27708}
\address{Jefferson Laboratory, 12000 Jefferson Ave. Newport News VA 23606}

\date{\today}

\maketitle

{\tighten
\begin{abstract}

The leading particle effect in charm hadroproduction is an enhancement of the cross section for a
charmed hadron $D$ in the forward direction of the beam when the beam hadron has a valence parton
in common with the $D$. The large $D^+/D^-$ asymmetry observed by the E791 experiment is an
example of this phenomenon.  We show that the heavy-quark recombination mechanism provides an
economical explanation for this effect.  In particular, the $D^+/D^-$ asymmetry can be fit
reasonably well using a single parameter whose value is consistent with a recent determination
from charm photoproduction. 

\end{abstract} } 

\vspace{0.33 in} ]\narrowtext

Fixed-target hadroproduction experiments have observed large asymmetries in the production of
charmed mesons and baryons\cite{Aitala:1996hf,Garcia:2001xj,Aitala:2000rd,Adamovich:1998mu}. These
asymmetries are commonly known as the ``leading particle effect'', since charmed hadrons having a
valence parton in common with the beam hadron are produced in greater numbers than other charmed
hadrons in the forward region of the beam. For example, the E791 experiment \cite{Aitala:1996hf}, in
which a $500~{\rm GeV}$ $\pi^-(\overline{u}d)$ beam is incident on a nuclear target, observes a
substantial excess of $D^-(\overline{c}d)$ over $D^+(c\overline{d})$ when the charmed mesons are
produced with  large momentum along the direction of the $\pi^-$ beam. The asymmetry,
\bea
\alpha[D^+] = {d\sigma[D^-]-d\sigma[D^+] \over d\sigma[D^-]+d\sigma[D^+]} \, ,
\eea
is as large as $\sim 0.7$ for the most forward $D$ mesons measured. Asymmetries in the
production of charmed baryons have also been observed
\cite{Garcia:2001xj,Aitala:2000rd,Adamovich:1998mu}.

In contrast with the large experimental asymmetries, perturbative QCD predicts
that the asymmetries should be very small.  The QCD factorization theorem for heavy particle
production \cite{Collins:1985gm} states that the cross section for producing a $D$ meson
in the collision of two hadrons can be written as
\bea\label{fact}
d\sigma &&[h h^\prime \rightarrow D + X] \\
&&=  \sum_{i,j} f_{i/h} \otimes f_{j/h^\prime} \otimes
d{\hat \sigma}(i j \rightarrow c \overline{c} +X) \otimes D_{c\rightarrow D} \, ,\nn
\eea
where $f_{i/h}$ is the distribution function for the parton  $i$ in the hadron $h$, $d{\hat \sigma}(i j
\rightarrow c \overline{c} +X)$ is the  parton cross section and $D_{c\rightarrow D}$ is the fragmentation
function for a $c$  quark hadronizing into a $D$ meson. The corrections to Eq.~(\ref{fact}) are suppressed
by powers of $\Lambda_{\rm QCD}/m_c$ or $\Lambda_{\rm QCD}/p_\perp$. A study of $O(\Lambda_{\rm QCD}/m_c)$
corrections to charm production can be found in Ref. \cite{Brodsky:1987cv}. The leading order (LO) diagrams
for $gg\rightarrow c\overline{c}$ and $q\overline{q} \rightarrow c\overline{c}$ produce $c$ and
$\overline{c}$ quarks symmetrically. The $c$ and $\overline{c}$ fragment independently into $D$ and
$\overline{D}$ mesons, and charge conjugation invariance requires that $D_{c\rightarrow D} =
D_{\overline{c}\rightarrow \overline{D}}$. Therefore, LO perturbative QCD predicts no asymmetry between $D$
and $\overline{D}$ mesons. Asymmetries are generated at next-to-leading order (NLO)
\cite{Nason:1989zy,Beenakker:1988bq,Beenakker:1990ma,Frixione:1994nb}, but they are too small by an order
of magnitude or more to account for the experimentally observed asymmetries.

Thus, charm asymmetries are interesting because they probe the power corrections to Eq.~(\ref{fact}).  Most
attempts to explain the leading particle effect have employed phenomenological models of hadronization. One
explanation is the ``beam drag effect''\cite{Norrbin:1998bw} in which charm quarks hadronize into charmed
mesons by the decay of a color string connected to partons in the  beam remnant. Calculations of the beam
drag effect employ the Lund string fragmentation model \cite{Sjostrand:1986ys}, as implemented in the
PYTHIA Monte Carlo\cite{Sjostrand:2001yu}. Predictions for the asymmetries are sensitive to the model for
the beam remnant, as well as other parameters in PYTHIA. These models can be tuned to fit data if one uses
a large charm quark mass and gives the partons intrinsic transverse momentum of $\sim 1\,{\rm
GeV}$\cite{Aitala:1996hf}. Another class of models  \cite{Vogt:1995fs} are based on the possibility of
intrinsic charm in the nucleon \cite{Brodsky:1980pb}. These models are sensitive to the poorly determined
intrinsic charm structure function as well as the model of the beam remnant and  predict smaller
asymmetries than are experimentally observed.

In this letter we show that the leading particle effect can be explained quantitatively and economically
by the heavy-quark recombination mechanism introduced in Ref.~\cite{Braaten:2001bf}. This mechanism has
already been applied to charm asymmetries in fixed-target photoproduction experiments in
Ref.~\cite{Braaten:2001uu}, where it was shown that the charm asymmetries observed in the experiments E687
and E691 at Fermilab can be fit well with just one free parameter. Here we apply the heavy-quark
recombination mechanism to the much larger asymmetries observed in the E791 hadroproduction experiment.

\begin{figure}[!t]
  \centerline{\epsfysize=4.5truecm \epsfbox[100 550 350 700]{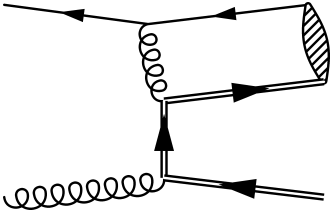}  }
 {\tighten
\caption[1]{Example of a diagram for $c\overline{q}$ recombination into a $D$ meson. Single lines are 
light quarks, double lines are heavy quarks and the shaded blob is the $D$ meson.}
\label{lo} }
\end{figure}

In the heavy-quark recombination mechanism, a heavy quark or antiquark recombines with a light
parton that participates in the hard-scattering process rather than a spectator parton from the
beam or target. A nonperturbative parameter that characterizes the probability for the light
parton to bind to the heavy quark to form a meson sets the overall  normalization of the cross
section. Since the asymmetry is generated in the hard process, the dependence of the asymmetry on
kinematic variables is calculable within perturbative QCD. Therefore, heavy-quark recombination
provides a simpler and more predictive explanation for the asymmetries than conventional
hadronization models.

An example of a Feynman diagram contributing to this process is depicted in Fig.~\ref{lo}.
A light antiquark $\overline{q}$ participates in a hard-scattering process which produces
a $c$ and  $\overline{c}$. The outgoing $\overline{q}$ emerges from the hard-scattering process
with momentum that is soft in the rest frame of the outgoing $c$ quark,
and the $c$ and $\overline{q}$ then bind to form a $D$ meson. There is an analogous
process in which a light quark combines with a $\overline{c}$ to form a $\overline{D}$ meson.
We emphasize that heavy-quark recombination is not taken into account by higher order
perturbative QCD corrections to the fragmentation contribution. The NLO correction  includes a parton
subprocess $\overline{q}g\rightarrow \overline{q}c\overline{c}$ that is similar to the parton
subprocess of the diagram in Fig.~\ref{lo}. However, for most of the phase space the outgoing
$\overline{q}$ hadronizes into a jet that is distinct from the jets containing the $c$ and
$\overline{c}$. When the $\overline{q}$ has small momentum in the $c$ rest frame, nonperturbative
effects can bind the $c$ and $\overline{q}$ and enhance the cross section. This enhancement, which is
not present in the NLO perturbative calculation, is accounted for by heavy-quark recombination.

The heavy-quark recombination contribution to the $D$ meson cross section is:
\bea
d{\hat\sigma}[D] = d{\hat\sigma}[\overline{q} g\rightarrow(c\overline{q})^n + \overline{c}]\,
\rho[(c \overline{q})^n \rightarrow D] \, .
\eea
This cross section must be convolved with parton distributions for the $\overline{q}$ and $g$ in the
colliding hadrons. The symbol $(c\overline{q})^n$ indicates that the $\overline{q}$ has momentum
$O(\Lambda_{\rm QCD})$ in the rest frame of the $c$ and that the $(c\overline{q})^n$ has the color and
angular momentum quantum numbers specified by $n$.  $d{\hat\sigma}(\overline{q}
g\rightarrow(c\overline{q})^n+\overline{c})$ is calculable in perturbative QCD, while  $\rho[(c
\overline{q})^n \rightarrow D]$ is a nonperturbative factor proportional to the probability for the
$(c\overline{q})^n$ to evolve into a state including the $D$ meson.

A detailed description of the calculation of $d{\hat\sigma}[\overline{q}
g\rightarrow(c\overline{q})^n+\overline{c}]$ can be found in Ref.~\cite{Braaten:2001bf}.   In this
letter, we  summarize the most important features. In most regions of phase space, the heavy quark
recombination contribution is power-suppressed relative to the fragmentation contribution in
Eq.~(\ref{fact}) in accord with the factorization theorems of QCD. Let $\theta$ be the angle between
the incoming $\overline{q}$ and the outgoing $D$ meson in the parton center-of-momentum frame. At
$\theta=\pi/2$, the ratio of the parton cross sections for heavy-quark recombination and LO
gluon-gluon fusion is
\begin{eqnarray}\label{sup}
\left.
{d \hat{\sigma}[\overline{q}g \rightarrow (c\overline{q})(^1S^{(1)}_0) +\overline{c}]  \over
d \hat{\sigma}[g g\rightarrow \overline{c} c]}\right|_{\theta = \pi/2}
&\approx &\, {256\,\pi\, \over 567} \alpha_s {m_c^2 \over p_\perp^2+m_c^2},  \nonumber \\
\nonumber \\
\left. {d \hat{\sigma}[\overline{q}g  \rightarrow (c\overline{q})(^3S^{(1)}_1) +\overline{c}]
\over d \hat{\sigma}[g g\rightarrow \overline{c} c]}\right|_{\theta = \pi/2}
&\approx &\, {256\,\pi\, \over 189} \alpha_s {m_c^2 \over p_\perp^2+m_c^2}. \nn
\end{eqnarray}
Using the expected scaling $\rho \sim \Lambda_{\rm
QCD}/m_c$, we find that heavy-quark recombination is an $O(\Lambda_{\rm QCD} m_c/p_\perp^2)$ power
correction for $p_\perp \gg m_c$.
However, the heavy-quark recombination cross section is strongly peaked when the $D$ is
produced in the forward region of the incoming $\overline{q}$. At $\theta = 0$,
the ratio of the parton cross sections is
\begin{eqnarray}
\left.
{d \hat{\sigma}[\overline{q}g \rightarrow (c\overline{q})(^1\!S^{(1)}_0,^3\!S^{(1)}_1) +
\overline{c}]  \over d \hat{\sigma}[g g \rightarrow \overline{c} c]}\right|_{\theta = 0}
\approx{256\,\pi\, \over 81} \alpha_s. \nonumber
\end{eqnarray}
Thus, there is no kinematic suppression factor  when the $(c\overline{q})^n$ is produced in the
forward region of the incoming $\overline{q}$. Heavy-quark recombination provides a natural
explanation for the leading particle effect because in the forward region, the cross section for a
$D(\overline{D})$ meson carrying a $\overline{q}(q)$ that matches a valence parton in the beam is
larger than that of other charmed mesons since valence quark structure functions are larger
than sea quark structure functions.

If the $(c\overline{q})^n$ hadronizes into a $D$ meson and nothing else, the $(c\overline{q})^n$
must be in a color-singlet state with the same flavor and angular momentum quantum numbers as the
$D$. This contribution to $\rho[(c\overline{q})^n \rightarrow D]$
is proportional to the square of a moment of the
$D$ meson light-cone wavefunction and it scales as  $\Lambda_{\rm QCD}/m_c$.
However one should also allow for nonperturbative transitions in which the $(c\overline{q})^n$
hadronizes into states including additional soft hadrons in the rest frame of the $D$.
The inclusive parameter $\rho[(c\overline{q})^n \rightarrow D]$, which includes
these additional contributions, is still expected to scale like $\Lambda_{\rm QCD}/m_c$.
Furthermore, the $(c\overline{q})^n$ can have
different color and angular momentum quantum numbers than the $D$ meson in the final
state.  Since the momentum of the light quark is $O(\Lambda_{\rm
QCD})$, amplitudes for production of $(c\overline{q})^n$ with $L>0$ are suppressed relative to
S-waves by powers of $\Lambda_{\rm QCD}/m_c$ or $\Lambda_{\rm QCD}/p_\perp$. The heavy-quark
recombination contribution to $D$ meson production thus contains four free parameters:
\bea\label{para}
\rho_1^{\rm sm} = \rho[c\overline{d}(^1S_0^{(1)})\rightarrow D^+],
\rho_1^{\rm sf} = \rho[c\overline{d}(^3S_1^{(1)}) \rightarrow D^+], \\
\rho_8^{\rm sm} = \rho[c\overline{d}(^1S_0^{(8)})\rightarrow D^+],
\rho_8^{\rm sf} = \rho[c\overline{d}(^3S_1^{(8)}) \rightarrow D^+]. \nn
\eea
Here the superscript $\rm sm$ stands for spin-matched and $\rm sf$ stands for spin-flipped.
(Nonperturbative transitions in which the light quark flavor of the $(c\overline{q})^n$ differs from
that of the $D$ are suppressed in the large $N_c$ limit of QCD and will be ignored in this analysis.)
Heavy-quark symmetry relates the four parameters appearing in the production of $D^{*+}$ to the four
parameters in Eq.~(\ref{para}). Parameters for $D^-$ and $D^0$ production can be related using charge
conjugation and isospin symmetry. The color-octet contributions are analogous to the color-octet
production matrix elements that play a prominent role in the theory of quarkonium
production\cite{Bodwin:1994jh}. If the $\overline{d}$ appearing in Eq.~(\ref{para}) were a heavy
antiquark, the parameters $\rho_1^{\rm sf},\rho_8^{\rm sm}$ and $\rho_8^{\rm sf}$ would be suppressed by
powers of $v$, where $v$ is the typical velocity of the antiquark in the bound state. However, the
$\overline{d}$ is light, so $v\sim 1$ and there is no apparent suppression of these parameters.

In photoproduction, the $O(\alpha \,\alpha_s^2)$ color-octet and color-singlet heavy-quark
recombination cross sections have the same functional form, so the cross section depends only on linear
combinations of color-singlet and color-octet parameters of the form $\rho_1+\rho_8/8$. The best fit to
fixed-target photoproduction asymmetries in Ref.~\cite{Braaten:2001uu} yielded $\rho_1^{\rm
sm}+\rho_8^{\rm sm}/8 = 0.15$. Including the $\rho^{\rm sf}$ parameters does not improve the fit unless
$\rho_1^{\rm sf}+\rho_8^{\rm sf}/8$ is negative, which is unphysical. In the case of fixed target
hadroproduction, the color-octet and color-singlet cross sections have different functional forms, so
the asymmetries depend on all four parameters. However, we find that the asymmetries measured by the
E791 experiment can be fit rather well by the single parameter $\rho_1^{\rm sm}$. In this analysis,
we only include $\rho_1^{\rm sm}$ and set the other parameters to zero.

There are two contributions to the heavy-quark recombination cross section for a $D$ meson:
\bea
&a)& \quad \sum_n d{\hat\sigma}[\overline{q} g\rightarrow(c\overline{q})^n+\overline{c}]\,
\rho[(c \overline{q})^n \rightarrow D] ,\\
&b)& \quad \sum_{q,n} d{\hat \sigma}[q g\rightarrow (\overline{c}q)^n+c]\,
\rho_n[(\overline{c}q)^n \rightarrow \overline{D}] \otimes D_{c \rightarrow D}.\nn
\eea
In process $a)$, a light antiquark recombines with a $c$ to form a $D$ meson. In process $b)$, a
$\overline{c}$ participates in the recombination process, and a $D$ meson is produced via
fragmentation from the $c$ quark which does not recombine. For $D^\pm$ production, process $b)$ 
partially dilutes the asymmetry generated by $a)$. We take $m_c=1.5 \,{\rm GeV}$ and set the
renormalization and factorization scales to be $\sqrt{p_\perp^2+m_c^2}$. The parton distributions are
GRV 98 LO \cite{Gluck:1998xa} for the nucleon and GRV-P LO \cite{Gluck:1991ey} for the pion. The E791
experiment uses a target consisting mostly of carbon, so the nucleon structure function is an isospin
symmetric linear combination of proton and neutron.  We use the one-loop expression for $\alpha_s$
with 4 active flavors and $\Lambda_{\rm QCD}= 200\,{\rm MeV}$. We include only the LO fragmentation
diagrams. If the NLO corrections are approximated  by including  a K factor, then $\rho_1^{\rm sm}$
should be multiplied by the same K factor to obtain the same asymmetry. We use the Petersen
parametrization for $D_{c\rightarrow D}$ with $\epsilon =0.06$ and the fragmentation probabilities are
determined from data in Ref. \cite{Gladilin:1999pj}. Contributions to the $D$ cross section from
feeddown from $D^*$ decay are included. Feeddown from other excited $D$ meson states is expected to be
small.

\begin{figure}[!t]
  \centerline{\epsfysize=8.5truecm \begin{turn}{270} \epsfbox{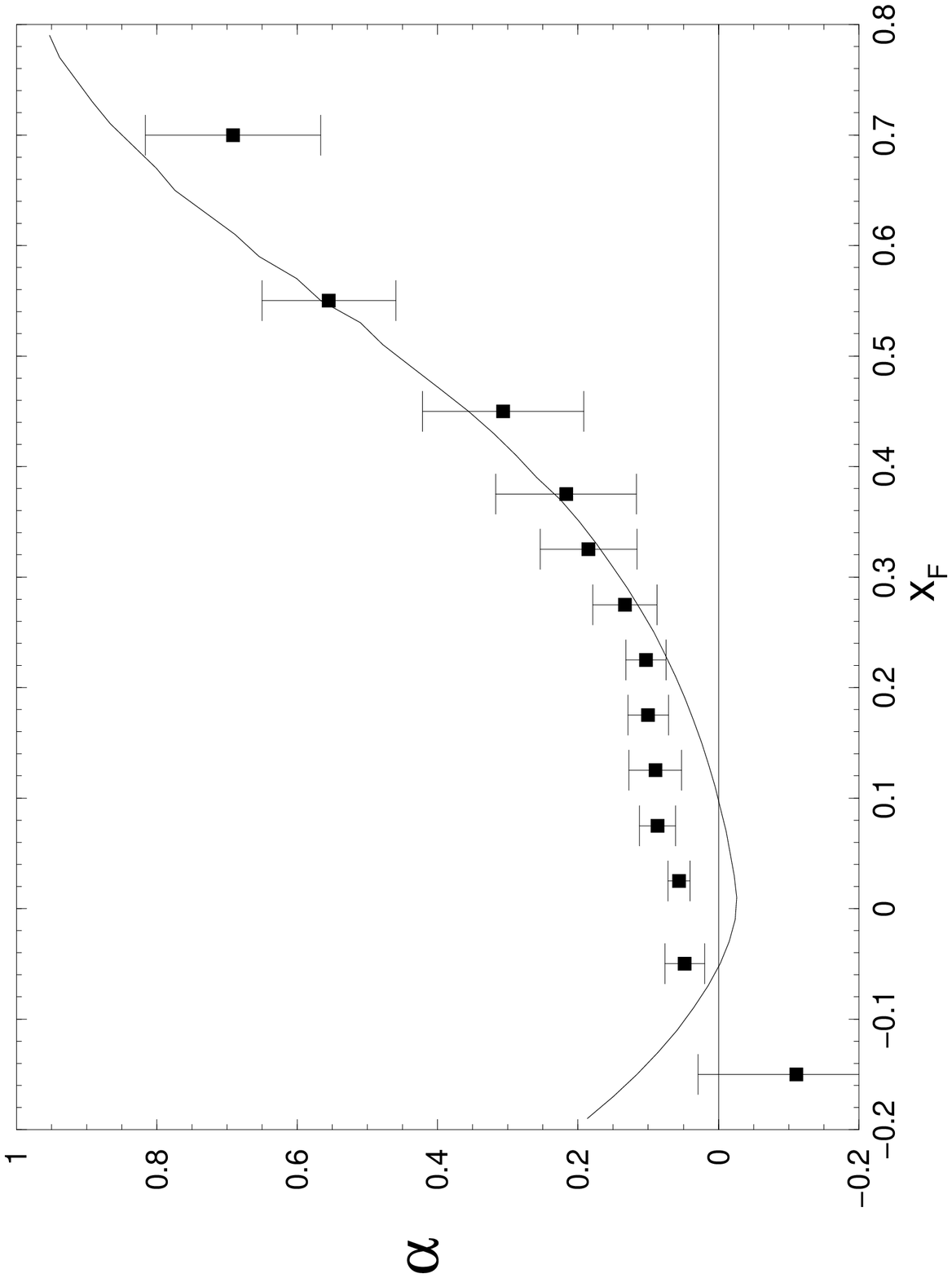}
  \end{turn} }
  \centerline{\epsfysize=8.5truecm \begin{turn}{270} \epsfbox{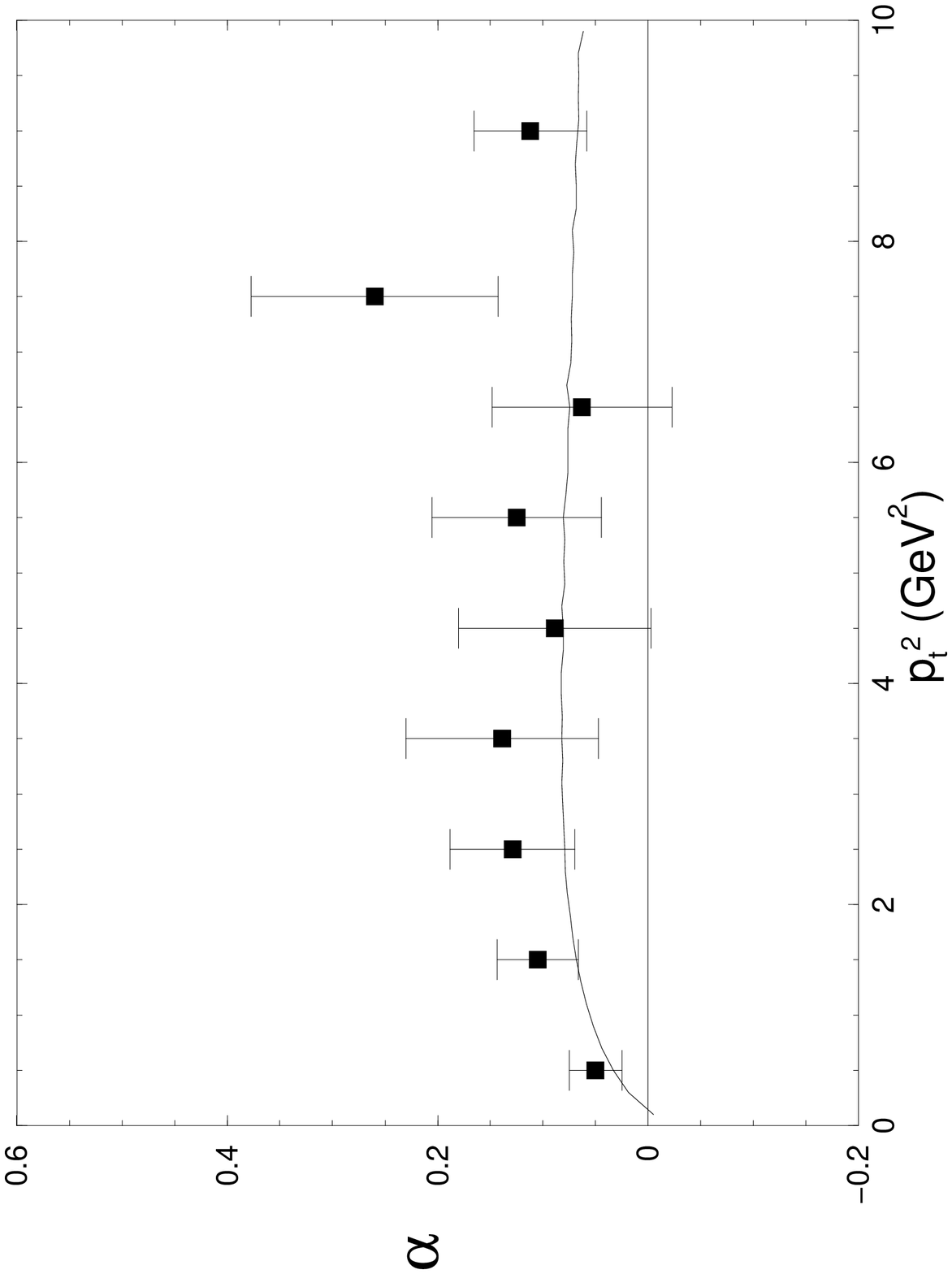}
  \end{turn} }
  \vspace{0.25 in}
 {\tighten
\caption[1]{
The asymmetry variable $\alpha[D^+]$ vs. $x_F$ and $p_\perp^2$. The data points are the measurements of 
E791 and the solid line is our prediction with $\rho^{\rm sm}_1 = 0.06$. }
\label{asym}}
\end{figure}

The predictions for $\alpha[D^+]$ as a function of $x_F$ and $p_\perp^2$ are compared to the E791 data in
Fig.~\ref{asym}. A least squares fit to all the data yields  $\rho^{\rm sm}_1=0.06$.  The $p_\perp$ dependence
is described very well by the heavy-quark recombination mechanism. The $x_F$ distribution is described well
for $x_F>0.2$, while our calculation underpredicts the asymmetry in the region $0.0 < x_F < 0.2$. It is
possible to obtain a better fit to the asymmetry in this region if one allows nonvanishing values of
$\rho_1^{\rm sf}$, $\rho_8^{\rm sm}$ and $\rho_8^{\rm sf}$. While we expect heavy-quark recombination to
dominate in the forward region, other $O(\Lambda_{\rm QCD}/m_c)$ corrections to fragmentation may be important
near $x_F = 0$ and could also account for the discrepancy. A systematic analysis of all $O(\Lambda_{\rm
QCD}/m_c)$ corrections to $D$ cross sections is needed to address this issue, but this is beyond the scope of
this paper. Although $\alpha$ was not measured at large negative $x_F$, it must be positive in this region in
accord with our prediction because of the leading particle effect associated with the target nucleons.

The value of $\rho_1^{\rm sm}$ extracted from the fits is consistent with the scaling $\rho_1^{\rm sm}
\sim O(\Lambda_{\rm QCD}/m_c)$.  A lower bound on $\rho_1^{\rm sm} $ can be obtained by assuming that
$\rho_1^{\rm sm}[(c\overline{q})^n\rightarrow D]$ is saturated by a final state consisting of only the
$D$. Then $\rho_1^{\rm sm}$ is proportional to the square of a moment  of the $D$ light-cone
wavefunction which can be bounded using heavy-quark effective theory arguments\cite{Korchemsky:2000qb}.
This value is smaller than the lower bound $\rho_1^{\rm sm}= 0.15$ from photoproduction fits
\cite{Braaten:2001uu}.   However, the uncertainty in this extraction due to  higher order perturbative
corrections and $O(\Lambda_{\rm QCD}/m_c)$ nonperturbative corrections could easily be a factor of 2. 
Furthermore, the fits in Ref.~\cite{Braaten:2001uu} used E687 data that was not corrected for
efficiencies because of correlations between the measured asymmetries and the production model used in
the Monte Carlo for simulating the trigger\cite{Frabetti:1996vi}. Given all these uncertainties, we
regard the values of $\rho_1^{\rm sm}$ extracted from hadroproduction and photoproduction to be
consistent.

It would be interesting to extend the analysis of this paper to asymmetries measured in
experiments with different beams and for different charmed hadrons, particularly charmed baryons.
Baryon asymmetries could be generated by a process similar to that of Fig.~\ref{lo} with the light
antiquark replaced by a light quark. There is an attractive force between the $c$ and $q$ when
their color state is the $\overline{3}$ of $SU(3)$ that can enhance the cross section in this
channel. It is natural to expect the $(cq)^n$ diquark to hadronize to a state with a charmed
baryon. The SELEX collaboration has measured $\Lambda_c^{\pm}$ asymmetries in experiments with
$\pi^-, \Sigma^-$ and $p$ beams \cite{Garcia:2001xj}. The $\Lambda_c^+$ and $\Lambda_c^-$ both 
have a valence quark that matches one of the valence quarks of the $\pi^-$, while only the
$\Lambda_c^+$ has a valence quark in common with the $p$ and $\Sigma^-$. Therefore, heavy-quark
recombination  predicts that $\Lambda_c^{\pm}$ asymmetries are much greater for $p$ and $\Sigma^-$
beams than for a $\pi^-$ beam. This prediction is in agreement with SELEX measurements, which find
$\alpha[\Lambda_c^-] \approx 1$ for $p$ and $\Sigma^-$ beams and $\alpha[\Lambda_c^-]\approx 0.2$
for $\pi^-$ beams.

We thank J. Appel for comments on this paper. E.B. and Y.J. are supported in part by Department of
Energy grant DE-FG02-91-ER4069. T.M. is supported in part by DOE grants DE-FG02-96ER40945 and
DE-AC05-84ER40150. T.M also acknowledges the hospitality of the University  of California at San
Diego Theory Group.


\begin{references}

%\cite{Aitala:1996hf}
\bibitem{Aitala:1996hf}
E.~M.~Aitala {\it et al.}  [E791 Collaboration],
Phys.\ Lett.\ B {\bf 371}, 157 (1996) ; {\it ibid}. {\bf 411}, 230 (1997);
G.~A.~Alves {\it et al.}  [E769 Collaboration],
Phys.\ Rev.\ Lett.\ {\bf 72}, 812 (1994); {\it ibid}. {\bf 77}, 2392 (1996);
%%CITATION = PHRVA,D49,4317;%%
M.~Adamovich {\it et al.}  [BEATRICE Collaboration],
Nucl.\ Phys.\ B {\bf 495}, 3 (1997);
M.~Adamovich {\it et al.}  [WA82 Collaboration],
Phys.\ Lett.\ B {\bf 305}, 402 (1993).

%\cite{Garcia:2001xj}<br>
\bibitem{Garcia:2001xj}
F.~G.~Garcia {\it et al.} [SELEX Collaboration],
%``Hadronic production of Lambda/c from 600-GeV/c pi-, Sigma- and p beams,''
Phys.\ Lett.\ B {\bf 528}, 49 (2002).
%%[arXiv:hep-ex/0109017].<br>
%%CITATION = HEP-EX 0109017;%%

%\cite{Aitala:2000rd}
\bibitem{Aitala:2000rd}
E.~M.~Aitala {\it et al.} [E791 Collaboration],
%``Asymmetries in the production of Lambda/c+ and Lambda/c- baryons in 500-GeV/c pi- nucleon interactions,''
Phys.\ Lett.\ B {\bf 495}, 42 (2000).
%[arXiv:hep-ex/0008029].
%%CITATION = HEP-EX 0008029;%%

%\cite{Adamovich:1998mu}
\bibitem{Adamovich:1998mu}
M.~I.~Adamovich {\it et al.} [WA89 Collaboration],
%``Charge asymmetries for D, D/s and Lambda/c production in Sigma-&nbsp; nucleus interactions at 340-GeV/c,''
Eur.\ Phys.\ J.\ C {\bf 8}, 593 (1999).
%[arXiv:hep-ex/9803021].
%%CITATION = HEP-EX 9803021;%%

%\cite{Collins:1985gm}
\bibitem{Collins:1985gm}
J.~C.~Collins, D.~E.~Soper and G.~Sterman,
%``Heavy Particle Production In High-Energy Hadron Collisions,''
Nucl.\ Phys.\ B {\bf 263}, 37 (1986).

%\cite{Brodsky:1987cv}
\bibitem{Brodsky:1987cv}
S.~J.~Brodsky, J.~F.~Gunion and D.~E.~Soper,
%``The Physics Of Heavy Quark Production In Quantum Chromodynamics,''
Phys.\ Rev.\ D {\bf 36}, 2710 (1987).
%%CITATION = PHRVA,D36,2710;%%

%\cite{Nason:1989zy}
\bibitem{Nason:1989zy}
P.~Nason, S.~Dawson and R.~K.~Ellis,
%``The One Particle Inclusive Differential Cross-Section For Heavy Quark Production In Hadronic Collisions,''
Nucl.\ Phys.\ B {\bf 327}, 49 (1989)
[Erratum-ibid.\ B {\bf 335}, 260 (1989)].

%\cite{Beenakker:1988bq}
\bibitem{Beenakker:1988bq}
W.~Beenakker, H.~Kuijf, W.~L.~van Neerven and J.~Smith,
%``QCD Corrections To Heavy Quark Production In P Anti-P Collisions,''
Phys.\ Rev.\ D {\bf 40}, 54 (1989).
%%CITATION = PHRVA,D40,54;%%

%\cite{Beenakker:1990ma}
\bibitem{Beenakker:1990ma}
W.~Beenakker, W.~L.~van Neerven, R.~Meng, G.~A.~Schuler and J.~Smith,
%``QCD Corrections To Heavy Quark Production In Hadron-Hadron Collisions,''
Nucl.\ Phys.\ B {\bf 351}, 507 (1991).
%%CITATION = NUPHA,B351,507;%%

%\cite{Frixione:1994nb}
\bibitem{Frixione:1994nb}
S.~Frixione, M.~L.~Mangano, P.~Nason and G.~Ridolfi,
%``Charm and bottom production: Theoretical results versus experimental data,''
Nucl.\ Phys.\ B {\bf 431}, 453 (1994).
%%CITATION = NUPHA,B431,453;%%

%\cite{Norrbin:1998bw}
\bibitem{Norrbin:1998bw}
E.~Norrbin and T.~Sj\"{o}strand,
%``Production mechanisms of charm hadrons in the string model,''
Phys.\ Lett.\ B {\bf 442}, 407 (1998).
%%[arXiv:hep-ph/9809266].
%%CITATION = HEP-PH 9809266;%%

%\cite{Sjostrand:1986ys}
\bibitem{Sjostrand:1986ys}
H.-U.~Bengtsson and T.~Sj\"{o}strand, Comput.\ Phys.\ Commun.\ {\bf 46}, 43 (1987).

%\cite{Sjostrand:2001yu}
\bibitem{Sjostrand:2001yu}
T.~Sjostrand, L.~Lonnblad and S.~Mrenna,
%``PYTHIA 6.2: Physics and manual,''
hep-ph/0108264.
%%CITATION = HEP-PH 0108264;%%

%\cite{Vogt:1995fs}
\bibitem{Vogt:1995fs}
R.~Vogt and S.~J.~Brodsky,
%``Charmed Hadron Asymmetries in the Intrinsic Charm Coalescence Model,''
Nucl.\ Phys.\ B {\bf 478}, 311 (1996).
%[arXiv:hep-ph/9512300].
%%CITATION = HEP-PH 9512300;%%

%\cite{Brodsky:1980pb}
\bibitem{Brodsky:1980pb}
S.~J.~Brodsky, P.~Hoyer, C.~Peterson and N.~Sakai,
%``The Intrinsic Charm Of The Proton,''
Phys.\ Lett.\ B {\bf 93}, 451 (1980).
%%CITATION = PHLTA,B93,451;%%

%\cite{Braaten:2001bf}
\bibitem{Braaten:2001bf}
E.~Braaten, Y.~Jia and T.~Mehen,
%``B production asymmetries in perturbative QCD,''
hep-ph/0108201.
%%CITATION = HEP-PH 0108201;%%

%\cite{Braaten:2001uu}
\bibitem{Braaten:2001uu}
E.~Braaten, Y.~Jia and T.~Mehen,
%``Charm anticharm asymmetries in photoproduction from perturbative recombination,''
hep-ph/0111296.
%%CITATION = HEP-PH 0111296;%%

%\cite{Bodwin:1994jh}
\bibitem{Bodwin:1994jh}
G.~T.~Bodwin, E.~Braaten and G.~P.~Lepage,
%``Rigorous QCD analysis of inclusive annihilation and production of heavy quarkonium,''
Phys.\ Rev.\ D {\bf 51}, 1125 (1995)
[Erratum-ibid.\ D {\bf 55}, 5853 (1995)].
%[arXiv:hep-ph/9407339].
%%CITATION = HEP-PH 9407339;%%

%\cite{Gluck:1998xa}
\bibitem{Gluck:1998xa}
M.~Gluck, E.~Reya and A.~Vogt,
%``Dynamical parton distributions revisited,''
Eur.\ Phys.\ J.\ C {\bf 5}, 461 (1998).
%[arXiv:hep-ph/9806404].
%%CITATION = HEP-PH 9806404;%%

%\cite{Gluck:1991ey}
\bibitem{Gluck:1991ey}
M.~Gluck, E.~Reya and A.~Vogt,
%``Pionic parton distributions,''
Z.\ Phys.\ C {\bf 53}, 651 (1992).
%%CITATION = ZEPYA,C53,651;%%

%\cite{Gladilin:1999pj}
\bibitem{Gladilin:1999pj}
L.~Gladilin, hep-ex/9912064.
%%CITATION = HEP_EX 9912064;%%

%\cite{Ackerstaff:1998as}
\bibitem{Ackerstaff:1998as}
K.~Ackerstaff {\it et al.}  [OPAL Collaboration],
%``Determination of the production rate of D*0 mesons and of the ratio  V/(V+P) in Z0 $\to$ c anti-c decays,''
Eur.\ Phys.\ J.\ C {\bf 5}, 1 (1998).
%[arXiv:hep-ex/9802008].
%%CITATION = HEP-EX 9802008;%%

%\cite{Korchemsky:2000qb}
\bibitem{Korchemsky:2000qb}
G.~P.~Korchemsky, D.~Pirjol and T.~Yan,
%``Radiative leptonic decays of $B$ mesons in QCD,''
Phys.\ Rev.\ D {\bf 61}, 114510 (2000).
%[hep-ph/9911427].
%%CITATION = HEP-PH 9911427;%%

%\cite{Frabetti:1996vi}
\bibitem{Frabetti:1996vi}
P.~L.~Frabetti {\it et al.} [E687 Collaboration],
%``Charm - anti-charm asymmetries in high-energy photoproduction,''
Phys.\ Lett.\ B {\bf 370}, 222 (1996).
%%CITATION = PHLTA,B370,222;%%


\end{references}
\end{document}